# Mass and Environment as Drivers of Galaxy Evolution III:

# The constancy of the faint-end slope and the merging of galaxies

Ying-jie Peng[1], Simon J. Lilly[1], Alvio Renzini[2], Marcella Carollo[1]

1. Institute of Astronomy, ETH Zurich, 8093 Zurich, Switzerland
2. INAF - Osservatorio Astronomico di Padova, Vicolo dell'Osservatorio 5, 35122, Padova, Italy

**ABSTRACT**

We explore using our continuity approach the underlying connections between the evolution of the faint end slope $\alpha_s$ of the stellar mass function of star-forming galaxies, the logarithmic slope $\beta$ of the sSFR-mass relation and the merging of galaxies. We derive analytically the consequences of the observed constancy of $\alpha_s$ since redshifts of at least $z \sim 2$. If the logarithmic slope $\beta$ of the sSFR-mass relation is negative, then the faint end slope $\alpha_s$ should quickly diverge due to the differential mass increase of galaxies on the star-forming main sequence and this will also quickly destroy the Schechter form of the mass function. This problem can be solved by removing low mass galaxies by merging them into more massive galaxies. We quantify this process by introducing the specific merger mass rate (sMMR) as the specific rate of mass added to a given galaxy through mergers. For a modest negative value of $\beta \sim -0.1$, an average sMMR $\sim 0.1$sSFR across the population is required to keep $\alpha_s$ constant with epoch, as observed. This in turn implies a merger rate of $\sim 0.2$sSFR for major mergers, which is consistent with the available observational estimates. More negative values of $\beta$ require higher sMMR and higher merger rates, and the steepening of the mass function becomes impossible to control for $\beta < -(\alpha_s+2)$. The close link that is required between the in situ sSFR and the sMMR probably arises because both are closely linked to the buildup of dark matter haloes. These new findings further develop the formalism for the evolving galaxy population that we introduced earlier, and show how striking symmetries in the galaxy population can emerge as the result of deep links between the physical processes involved.

# 1 INTRODUCTION

Understanding the history of star formation and stellar mass assembly in the galaxy population is an important goal of cosmology. One of the central questions in a hierarchical cosmology is the relative importance of *in situ* star-formation and the accretion of pre-formed stars through the merging of smaller galaxies, in building up the stellar mass of typical galaxies. When we look at the stars within a given galaxy, we can ask what fraction of them were formed within that gravitational potential well and what fraction were formed outside, in another potential well that subsequently merged into the galaxy in question. Motivated by the existence of highly luminous star-bursts associated, at least in the local Universe, with mergers of galaxies, we could also identify a third category of stars, namely those that are actually formed *during* a merger event. However, we will not consider this third category further in this paper.

Over the last decade, it has become increasingly clear that there is a tight relation between the instantaneous star-formation rate (SFR) of most star-forming galaxies and their stellar masses $m$, at least up to $z \sim 2$, defining the so-called Main Sequence of star-forming galaxies (e.g. Daddi et al. 2007; Elbaz et al. 2007; Noeske et al. 2007; Peng et al 2010). The specific SFR (sSFR) is a power-law in the stellar mass of the galaxy, $sSFR \sim m^{\beta}$. The vast majority of star-forming galaxies lie on this Main Sequence. A minority of star-burst galaxies that may be the result of major mergers (Sanders & Mirabel 1996) have substantially elevated sSFRs and therefore lie above this relation. However, these star-burst galaxies are a minority and represent only a few percent of all star-forming galaxies, accounting for about 10% of the overall star-formation over the $0 < z < 2$ range (e.g. Rodighiero et al 2011). The fraction of such outliers appears to be more or less constant over the redshift interval $0 < z < 2$ (Sargent et al 2012). The existence of the galaxy Main Sequence suggests that the star-formation in most galaxies, and most of the star-formation in the Universe, is fueled mainly by smooth accretion of material (e.g. Dekel & Birnboim 2006; Daddi et al. 2007; Elmegreen et al. 2009; Dekel et al. 2009; Lilly et al 2013) rather than by cataclysmic major mergers of gas-rich galaxies producing short-lived but very vigorous episodes of star-formation.

However, this does not address the primary issue, which is how many pre-formed stars are brought into a galaxy relative to the number that are formed in situ within the galaxy. Observationally, determination of the merger rates of galaxies is quite difficult (e.g. Lotz et al 2011, Kampczyk et al 2013), mostly due to the difficulties of isolating merger events, either kinematically or morphologically, and uncertainties in estimating the timescales over which the merger will take place in each case which is required to go from an observed "fraction" to a "rate" per galaxy. An alternative approach is to look at the mass-function of galaxies and to infer the action of merging from changes in this, taking into account also the ongoing star-formation in the galaxies (e.g. Drory et al. 2008, Pozzetti et al. 2010, Moustakas et al. 2013). When directly tied to observational data, this can run into difficulties due to issues such as cosmic variance affecting the mass functions at closely spaced epochs.

In previous papers in this series (Peng et al 2010, 2012, hereafter P10 and P12 respectively) we have developed a formalism for considering the evolution of the galaxy population that is based on identifying a limited number of key simplicities in the overall galaxy population and then exploring, analytically, the consequences of these using the most basic continuity equations. In P10, we focused on the implications of the observed separability of the red fraction $f_{red}$ of galaxies in the SDSS and the observed constancy of the characteristic M* of the mass-function of star-forming galaxies since $z \sim 2$ (e.g., Ilbert et al. 2010). The red fraction of galaxies tells us, at a given mass and/or in a given environment, the fraction of galaxies that have been "quenched", i.e. in which star-formation has been suppressed by one to two orders of magnitude relative to the "blue" galaxies on the Main Sequence. In P10, we argued that the observed separability of $f_{\text{red}}$ in terms of the stellar mass and environment of galaxies indicates that there are two main channels of quenching, one dependent on environment, which we call "environment-quenching", and the other linked to the mass of the galaxy, or "mass-quenching". The observed constancy of M* of the star-forming population then places strong constraints on the form of mass-quenching, namely that the rate of mass-quenching must be given by, or at least closely mimic, the star-formation rate of a galaxy divided by the time-independent M*. Mass-quenching must be independent of environment and environment-quenching must be independent of stellar mass, from separability. We showed in P12, that environment-quenching is confined to "satellite" galaxies, i.e. galaxies that are in the halo of another dominant galaxy, the "central" galaxy.

One of the great successes of the formalism introduced in P10 is that it predicts a number of well-defined inter-relationships between the Schechter parameters describing different components of the galaxy population. If we look at the mass-functions of star-forming and passive galaxies in high and low density environments, or that are centrals or satellites, we find that there should be a common universal value of M*, set by the form of the mass-quenching process (as above). The faint-end slope of the mass-quenched passive galaxies should differ from that of the star-forming galaxies by $\Delta\alpha_s \sim 1$. The faint-end slope of the environment-quenched galaxies should be the same as that of the star-forming population. This should produce, in satellite galaxies, a characteristic double Schechter function for the overall population of passive galaxies. Finally, a double Schechter function is also predicted for the overall mass function of the entire galaxy population. To a very large degree these inter-relationships are observed in practice (see P10, P12, Baldry et al 2012). Most relevantly for the current paper, the predicted change in faint end slope is $\Delta\alpha_s \sim 1$, as indeed observed (P10, P12, Baldry et al 2012, Knobel et al. 2013).

In passing, and as noted in P10 and P12, there is some evidence for small deviations from the predicted relations for passive galaxies, especially at the massive end above M*. This was interpreted as the signature of the effects of post-quenching merging of galaxies, or, to be more precise, any mass increase after the galaxies leave the blue star-forming population. This leads to a small increase in the M* of passive central galaxies relative to the (constant) M* of star-forming central galaxies, and a small increase in M* and associated small decrease in $\alpha_s$ for the overall passive population (combining centrals and satellites) in high density environments. The size of this effect is however quite small, and we can estimate that the average mass-increase due to this post-quenching merging is about 35% for passive central galaxies. It should be noted that this is 10% higher than in the original P10 and P12 papers because in those we omitted the small additional effect due to the return of mass as stellar populations age.

In the P10 analytic formalism, the effects of merging were included through two $\kappa$ terms representing the flux of galaxies *out* of ($\kappa$-) and *into* ($\kappa$+) a given mass interval due to mergers. In P10, it was assumed that all mergers would lead to the quenching of the surviving galaxy. As a result, the $\kappa$- became another quenching channel that could be lumped into the general environment-quenching term. The $\kappa$+ was not dealt with explicitly, but would be associated with the change in M* of the passive population that was discussed in the previous paragraph. We will return to the issue of $\kappa$ in a later paper.

In this paper, we wish to look at the effect of mergers that do not lead to the quenching of the surviving galaxy. This was not included in the P10 formalism, although the effect on the star-forming survivor could, to a certain degree, be dealt with by adjusting the effective star-formation rate to account for the increase in mass. This is because the P10 continuity formalism was primarily concerned with the increase in stellar mass of a given galaxy, independent of whether the new stars were formed in situ or brought in to the galaxy via a (non-quenching) merger, and so adjusting the SFR upwards from the observed value therefore has the same effect, for the survivor, as explicitly including the mass brought in from a merger.

In this paper, however, we will explicitly differentiate between these different modes of stellar mass growth, referring to the increase in stellar mass due to (non-quenching) mergers as the merger mass rate (MMR), as opposed to the SFR from continuing *in situ* star-formation. Likewise, we will have the specific MMR (sMMR) as the equivalent to the sSFR. The relative sizes of the sMMR and the sSFR within the galaxy population therefore address the question raised at the beginning of this paper, namely the relative importance of mergers and *in situ* star-formation in building up the stellar populations of galaxies, and specifically of galaxies that are still on the star-forming Main Sequence.

As noted above, the analysis of P10 was heavily based on the remarkable fact that the observed M* of the population of star-forming galaxies is independent of epoch back to at least z ~ 2, and quite likely beyond, despite the large increase in the stellar masses of any galaxy that stays on the Main Sequence during this time. In this paper, we will instead focus on the second remarkable constancy exhibited by the mass function of star-forming galaxies: namely that the faint-end slope, $\alpha_s$, is also more or less constant (Ilbert et al. 2010). In P10, this was just taken as an observational fact - albeit an important one - as it enabled us to simply superpose the mass-functions of the passive galaxies that were produced from the star-forming population at different epochs, since they would always have the same M* (as above) and the same $\alpha_s$.

If β is identically zero, i.e. if all star-forming galaxies have the same sSFR, and if the number of galaxies is conserved, then the faint end slope $\alpha_s$ will stay constant, simply because all star-forming galaxies will increase in logarithmic stellar mass equally. However, it is clear that if the logarithmic slope β of the sSFR-mass relation is negative, then the $\alpha_s$ of the star-forming mass function should quickly steepen (if numbers are conserved) due to the differential mass increase of galaxies along the Main Sequence - lower mass galaxies will increase their logarithmic masses more rapidly than higher mass ones. Many recent studies have found a shallow negative slope of β ~ −0.1 (e.g. Elbaz et al. 2007; Daddi et al. 2007; Dunne et al. 2009; P10), although steeper values, with β ~ −0.4 have also been claimed (e.g. Noeske et al 2007, Karim et al 2011, Rodighiero et al 2010, although the last of these has now been revised to a much shallower β ~ −0.2 in Rodighiero et al 2011). The discrepancy in the observed values of β could be due to observation biases, the definition of the star-forming population, and/or different star-formation rate estimators (see e.g. Stringer et al. 2011; Whitaker et al. 2012; Salmi et al. 2012). As we will show in this paper, even small values of β ~ −0.1 will lead to a rapid steepening of the faint-end slope $\alpha_s$, making the observational result of a constant $\alpha_s$ to z ~ 2 (and, again, quite possibly to higher redshifts) all the more interesting.

The goal of this paper is to explore the consequences of the observed constant $\alpha_s$ in the same way that our first paper (P10) explored the consequences of constant M*. In particular, we will see that the destruction of galaxies through merging is an attractive way to keep $\alpha_s$ constant, especially for small values of β. *This requires a phenomenological connection between the merger rate and the cosmic* *evolution of the sSFR of Main Sequence galaxies that may, however, reflect a deep underlying connection between different aspects of structure formation in the Universe.*

The layout of the paper is as follows. In Section 2, we summarize the observations of $\alpha_s$ of the mass function of star-forming galaxies as a function of redshift. In Section 3 we derive an analytic expression for development of $\alpha_s(t)$ in terms of $\beta$. In Section 4 we introduce a way of self-consistently combining the sSFR, sMMR and merger-driven destruction of galaxies. This enables us, in Section 5, to calculate the amount of merging, and specifically the sMMR relative to the sSFR, that are required to keep $\alpha_s$ constant in the face of a non-zero $\beta$. We show that the required merger rate is consistent with the (uncertain) observational estimates. We discuss these findings in Section 6, and provide a summary of the paper in Section 7.

The cosmological model used in this paper is a concordance $\Lambda$CDM cosmology with $H_0 = 70$ kms$^{-1}$Mpc$^{-1}$, $\Omega_\Lambda = 0.75$ and $\Omega_M = 0.25$. Throughout, we use the term "dex" to mean the anti-logarithm, i.e. 0.1 dex = $10^{0.1}$ = 1.258.

## 2 MEASUREMENTS OF THE FAINT END SLOPE $\alpha_s$ OF THE STELLAR MASS FUNCTION OF STAR-FORMING GALAXIES

In Figure 1, we show measurements of the Schechter parameter $\alpha_s$ as determined from the observed stellar mass function of star-forming galaxies as function of cosmic time, taken from various sources (Pérez-González et al. 2008, Ilbert et al. 2010 & 2013, P10, González et al. 2011, Kajisawa et al. 2011, Lee et al. 2012, Baldry et al. 2012). At low redshift, it is clear that the mass-function of star-forming galaxies has $\alpha_s \sim -1.4$ (P10, Baldry et al 2012). The Baldry et al (2012) analysis combines the SDSS data with deeper GAMA spectroscopy extending the mass range present in P10. In passing, it should be noted that shallower faint end slopes are obtained if one force fits the overall mass function of galaxies (adding both star-forming and passive together) by a single Schechter function. However, this is because the overall mass function is both observed and expected (P10) to be a *double* Schechter function with the two components differing in $\alpha_s$ by approximately unity.

Most impressive is the star-forming mass-function from the homogeneous COSMOS data set of Ilbert et al (2010), re-fit as in P10 (Table 1) whose $\alpha_s$ values are close to $\alpha_s \sim -1.3$ with quite small error bars over the entire redshift range $0.2 < z < 2$. The older analysis of Perez-Gonzalez et al (2008) also has a constant $\alpha_s$ over an even larger redshift range $0.1 < z < 4$, but at a slightly lower value $\alpha_s \sim -1.2$. Several analyses of the mass function at $z > 2$ get the same value, albeit with large error bars (Gonzalez et al 2011, Lee et al 2012). Although there are systematic differences from sample to sample, $\alpha_s$ appears to be remarkably constant at least up to $z \sim 2$, and probably up to $z \sim 4$.

At even higher redshifts, $z > 4$, there may be a hint that the slope $\alpha_s$ is getting steeper but these measurements are highly uncertain since the completeness mass of a given survey progressively increases with increasing redshift, approaching M* (which may also be decreasing at these redshifts, see the discussion in P10). Moreover, it is still unclear whether the galaxy Main Sequence is already in place at $z > 4$ and thus the slope $\beta$ is essentially unconstrained at these redshifts. There is some indirect evidence that the slope $\alpha_s$ for the SFR functions show no clear evolution with cosmic time from $z \sim 0$ to $z \sim 7$ (Smit et al. 2012) which would be consistent with constant $\alpha_s$ and $\beta$. In the present work, we will mainly focus on the redshifts range where the $\alpha_s$ is mostly securely established to be more or less constant with time, i.e. $0 < z < 2$.

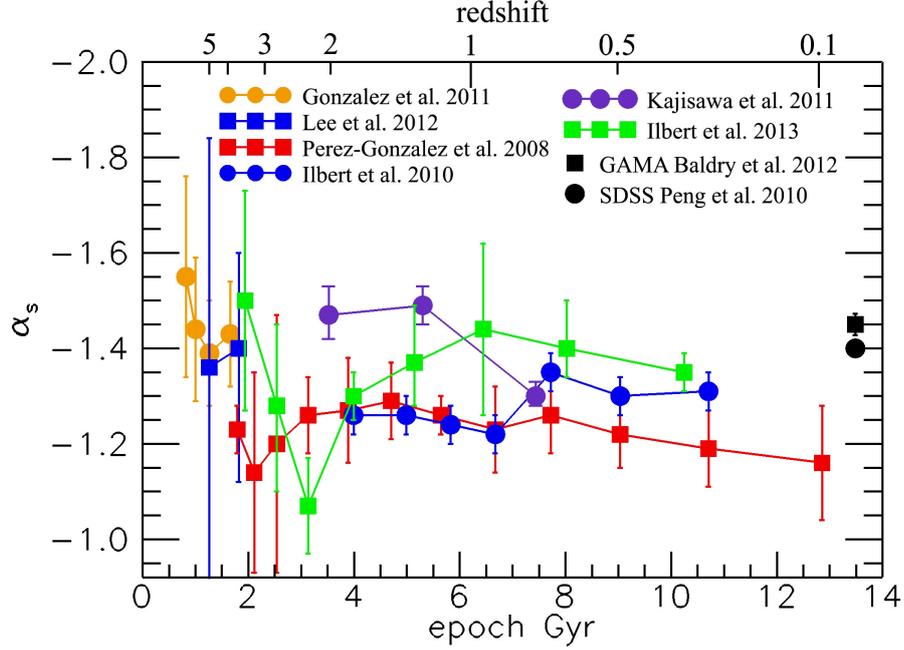

Figure 1. Evolution of the Schechter faint-end slope parameter $\alpha_s$ of the observed stellar mass function of star-forming galaxies with cosmic time from the literature. Although there are systematic differences from sample to sample the impression of constant $\alpha_s$ with epoch is strong within individual samples, and comparing the highest and lowest redshifts.

## 3 THE CONNECTION BETWEEN $\alpha_s$ AND $\beta$

### 3.1 The evolution of $\alpha_s$ for different $\beta$

The well-known Schechter function is parameterized by three quantities, the characteristic mass $M^*$, the faint-end slope $\alpha_s$ and the normalization $\phi^*$. The mass function is then given by

$$\phi dm = e^{-\frac{m}{M^*}} (\frac{m}{M^*})^{\alpha_s} \frac{\phi^*}{M^*} dm \quad (1)$$

The total number density of the population is given by integrating the mass function above some lower limit

$$N = \int_{M_{min}}^{\infty} \phi dm$$
$$= \phi^* \Gamma(\alpha_s + 1, \frac{M_{min}}{M^*}) \quad (2)$$

where $\Gamma$ is the gamma function defined as

$$\Gamma(x) = \int_0^{\infty} e^{-t} t^{x-1} dt \quad (3)$$

$M_{min}$ is the minimum stellar mass of the galaxy population and $\Gamma(\alpha_s+1, M_{min}/M^*)$ represents the integral of the gamma function from $M_{min}/M^*$ to infinity.

We can parameterize the sSFR(*m,t*) of star-forming galaxies on the Main Sequence as Equation (1) in P10 by

$$\text{sSFR}(m,t) = k(t)\left(\frac{m}{M^*}\right)^{\beta} \qquad (4)$$

where β is the logarithmic slope of the sSFR-mass relation and *k* is the value of *sSFR* at *m* = M*. As a detail, and as discussed in Lilly et al (2013), we want the sSFR to be the specific increase rate in stellar mass. The sSFR should therefore be computed using consistent definitions of the stellar mass, i.e. the SFR divided by the past integral of the SFR, rather than by the actual stellar mass present after the effects of mass return. This quantity, the *r*sSFR in Lilly et al (2013), will be smaller than the sSFR which is often quoted by observers, which is the instantaneous SFR divided by the actual, long-lived, stellar mass, by a factor (1-*R*), where *R* is the mass return fraction with R = 0.4 from stellar population models (e.g. Bruzual & Charlot 2003). Throughout this paper we will use the sSFR to denote the reduced specific star-formation rate, i.e. the inverse mass-doubling timescale of the long-lived stellar population. This was denoted rsSFR in Lilly et al (2013) and is, it should be noted, reduced by a factor of 0.6 from the sSFR considered in P10.

Clearly a negative value of β implies that low mass galaxies grow faster than more massive ones, since the sSFR$^{-1}$ is the e-fold mass increase timescale, or the logarithmic increase in stellar mass per unit time. If we conserve the numbers of star-forming galaxies, i.e. we neither quench nor merge any of the galaxies, then a negative β will lead to the steepening of the faint end slope of the star-forming mass function, as schematically illustrated in Figure 2.

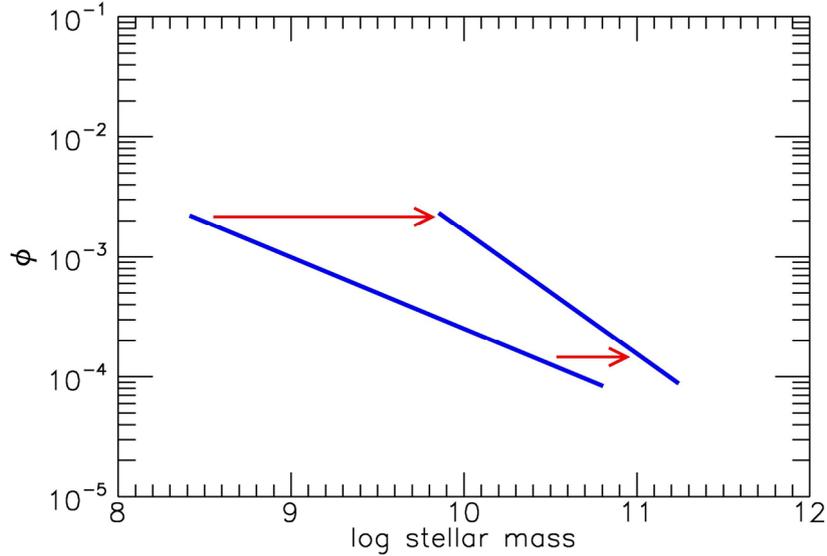

Figure 2. Schematic diagram to show the steepening of the faint end slope of the star-forming mass function that arises from the differential mass increase of galaxies for a negative slope β of the sSFR-*m* relation. The blue lines represent the power law part of the mass function of the star-forming galaxies and the red horizontal arrows represent the increase in mass (proportional to the sSFR), which decrease with increasing mass for a negative β.

The connection between α$_s$ and β can be quantified through continuity equations. Following Equation (20) in P10, and without environment-quenching, the mass function of the star-forming galaxies φ$_{blue}$(m,t) is given by:

$$\phi_{blue}(m,t) = \phi_{blue}(m,t_0) \cdot e^{\int_{t_0}^{t} -[1+\alpha_s(t)+\beta]\cdot \text{sSFR}(m,t)\, dt} \tag{5}$$

where φ$_{blue}$(m, t$_0$) is the mass function of the star-forming galaxies at some earlier time t$_0$.

Differentiating and rearranging Equation (5), we get

$$\frac{\partial \ln \phi_{blue}(m,t)}{\partial t} = -[1+\alpha_s(t)+\beta]\cdot \text{sSFR}(m,t) \tag{6}$$

At given stellar mass m, the time derivative of the local logarithmic slope of the mass function, which following P10 we designate as α(t), is given by

$$\frac{\partial \alpha(t)}{\partial t} = \frac{\partial}{\partial t}\left(\frac{\partial \log \phi_{blue}(m,t)}{\partial \log m}\right)$$
$$= \frac{\partial}{\partial \log m}\left(\frac{\partial \log \phi_{blue}(m,t)}{\partial t}\right) \tag{7}$$

Since α(t) = 1+α$_s$(t) − m/M*, we get (assuming constant M*, see P10) at a given m,

$$\left.\frac{\partial \alpha(t)}{\partial t}\right|_m = \left.\frac{\partial \alpha_s(t)}{\partial t}\right|_m$$

(8)

Inserting Equation (6) into (7) and considering Equation (8), we get, at given m,

$$\left.\frac{\partial \alpha_s(t)}{\partial t}\right|_m = -\text{sSFR}(m,t)\cdot \beta[1+\alpha_s(t)+\beta]\Big|_m \tag{9}$$

where we have used the simple facts that dlogφ = dlnφ / ln10 and dsSFR / dm = β sSFR/m from the definition of sSFR in Equation (4).

Solving the differential Equation (9) at given m, we find

$$\alpha_s(t) = [1+\alpha_s(t_0)+\beta]e^{-\int \text{sSFR}(m,t)\cdot \beta\, dt} - (1+\beta) \tag{10}$$

where α$_s$(t$_0$) is the α$_s$(t) at some earlier time t$_0$. We can rearrange the above equation into a more symmetrical form of:

$$1+\alpha_s(t)+\beta = [1+\alpha_s(t_0)+\beta]\chi^{-\beta} \tag{11}$$

where $\chi$ is the factor by which the stellar mass has increased due to star formation during the time interval between $t_0$ and $t$, and is given by

$$\chi = \frac{m(t)}{m(t_0)} = e^{\int_{t_0}^{t} \mathrm{sSFR}(m,t)\, dt} \tag{12}$$

It should be noted that $\chi$ will itself generally be a function of mass, unless $\beta = 0$, and this differential mass increase will not only steepen the mass function, but also destroy the pure Schechter form of the mass function. This can also be seen in Equation (11) above and is clearly shown in Figure 4.

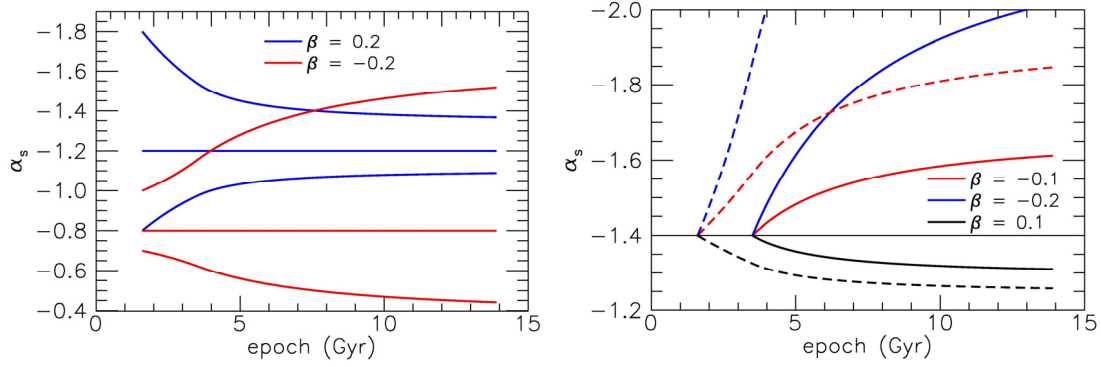

Figure 3. For non-zero $\beta$, the predicted evolution of the Schechter parameter $\alpha_s$ of the stellar mass function of the star-forming galaxies with cosmic time determined from Equation (10). In the left panel, we start the evolution at $z = 4$ with different initial values of $\alpha_s$ for $\beta = -0.2$ and $\beta = 0.2$, respectively. For negative $\beta$, $\alpha_s$ quickly diverges with time and for positive $\beta$, $\alpha_s$ converges towards $1+\beta = 1.2$. In the right panel, the solid lines show if we start the evolution with $\alpha_s = -1.4$ at $z=2$, the steepening (for negative $\beta$) or the flattening (for positive $\beta$) of the $\alpha_s$. The dashed lines show that the divergence of $\alpha_s$ is even faster if we start (with the same $\alpha_s$ and $\beta$) at an earlier epoch of $z=4$.

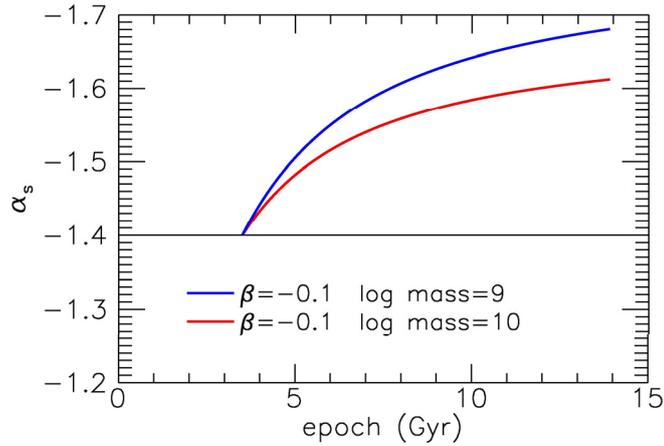

Figure 4. The differential steepening of $\alpha$ at two different stellar mass for $\beta = -0.1$ and the evolution starts with $\alpha_s = -1.4$ at $z = 2$. This differential divergence of $\alpha$ will quickly destroy the Schechter form of the mass function since the steepening rate, i.e. $d\alpha_s(t)/dt$, is different at different mass.

It is clear from Equation (11) that the $\chi^{-\beta}$ term controls the amplitude of the change to $\alpha_s$. If $\beta = 0$, $\alpha_s(t) = \alpha_s(t_0)$, i.e. $\alpha_s$ will be constant with time, as expected, for any value of $\alpha_s$. In this case the normalization of the mass function at given $m$ will increase with epoch if $\alpha_s < -1$, as can be seen in Equation (6) with $\beta = 0$. If $\beta > 0$ (which is not currently indicated by observational data) the $\chi^{-\beta}$ term in Equation (11) decreases towards zero with time, faster for larger $\beta$, and $\alpha_s$ will converge towards a value of $-(1+\beta)$, regardless of the initial value of $\alpha_s$.

For $\beta < 0$ (as indicated by observations) then it is clear from Equation (11) that a special case of $\alpha_s = -(1+\beta)$ can be maintained, but this is an unstable solution and moreover far from any observational determination of $\alpha_s$ at any redshift. Generally, $\alpha_s$ will diverge *away* from this value as the $\chi^{-\beta}$ term increases with time in Equation (11). The divergence will be faster for more negative values of $\beta$.

These different behaviors are illustrated in the two panels of Figure 3 where we have included the observed cosmological evolution of the reduced sSFR (at M*) in the Universe, reduced by a factor (1-R) = 0.6 from Equation (1) in P10. In the left hand panel, we show the convergence of $\alpha_s$ towards $-(1+\beta)$ for positive $\beta$ and the divergence away from $-(1+\beta)$ for negative $\beta$. The case $\beta = 0$ produces constant $\alpha_s$ for all values of $\alpha_s$. In the right hand panel we show the divergence for an initial $\alpha_s(t_0) = -1.4$ for three different values of $\beta$, and for $t_0$ set to the epoch corresponding to $z = 2$ and $z = 4$.

It is clear that either $\beta = 0$ (for any value of $\alpha_s$) or the special case of $\alpha_s = -(1+\beta)$ (for any value of $\beta$) can keep the faint end slope $\alpha_s$ constant with epoch (although the latter case is unstable for negative $\beta$). The important difference is that the case of $\alpha_s = -(1+\beta)$ also sets $d\ln\phi_{blue}/dt = 0$ at given $m$, i.e. the mass function (or number density) of the star-forming population at given $m$ will also be constant with time. For the $\beta = 0$ cases, the mass function at given $m$ will either increase with time (for $\alpha_s < -1$), decrease (for $\alpha_s > -1$) or stay the same (for $\alpha_s = -1$, i.e. for a flat mass function), as might be intuitively expected. Moreover, it is clear from Equation (9) that the time-derivative of $\alpha$, will always scale with the sSFR since this sets the e-fold time of the mass increase.

In practical terms, with the observed values of $-0.4 < \beta \leqslant 0$ and with $\alpha_s \sim -1.4$, we will have a situation in which differential mass growth will quickly steepen the $\alpha_s$, as in the red solid curve in the right hand panel of Figure 3. This makes the observed constancy of $\alpha_s$ since at least $z \sim 2$, and possibly earlier, quite remarkable, and strongly suggests that some other process must be at work.

One way to control the steepening of the mass function is to destroy low mass galaxies by merging them into larger ones. This both reduces the number density of low mass galaxies, and increases the mass of the larger galaxies. Both of these will act to counter the steepening effect and could contrive to keep the $\alpha_s$ constant over time. The main goal of this paper is to explore this possibility.

### 3.2 Another mass-dependent quenching term?

At the end of the previous section, we proposed that a mass dependent merging of the star-forming galaxies (without quenching) could counter the rapid steepening of the mass function by removing low mass galaxies from the mass-function and adding the associated stellar mass to higher mass star-forming galaxies. One might ask whether it is also then possible to achieve the same result simply by an additional mass-dependent quenching channel λ' that removes star-forming galaxies and turns them into passive ones.

Following Equation (19) in P10, or simply adding this additional quenching term λ' on the right hand side of Equation (6), the continuity equation of the star-forming galaxies at given mass with quenching is given by:

$$\frac{1}{\phi_{blue}(t)}\frac{\partial \phi_{blue}(t)}{\partial t} = -\text{sSFR}(t)\cdot(\alpha+\beta) - \lambda_m(t) - \lambda'(t)$$
$$= -\text{sSFR}(t)\cdot(1+\alpha_s+\beta) - \lambda'(t) \quad (13)$$

where $\lambda_m$ = SFR/M* is the mass-quenching rate, λ' is the proposed additional mass-dependent quenching rate and we have also used $\alpha=(1+\alpha_s)-m/M^*$. For a negative value of β, the term $-\text{sSFR}(1+\alpha_s+\beta)$ is positive (if $\alpha_s<-\beta-1$) and is proportional to the sSFR, i.e. to $m^\beta$. To keep $\alpha_s$ constant over cosmic time, it requires $d\log\phi_{blue}/dt$ to be independent of mass in the power law part, i.e. at $m < M^*$. From Equation (13), this requires $\lambda' = -\text{sSFR}(1+\alpha_s+\beta) + C(t)$, where $C(t)$ is some mass independent term. It thus seems possible at first sight to keep $\alpha_s$ constant with an additional mass dependent quenching rate.

However, if this new quenching term is mass dependent, i.e. $\lambda' \sim -\text{sSFR}(1+\alpha_s+\beta) \sim -0.5\text{sSFR}$ (for $\alpha_s \sim -1.4$ and $\beta \sim -0.1$), it will quench low mass star-forming galaxies more efficiently than the mass-quenching term $\lambda_m = \text{SFR}/M^* = \text{sSFR}\cdot m/M^*$ that we introduced in P10, and will dominate at low masses of $m < 0.5M^*$. This will produce a component of the passive mass function with slightly steeper (more negative) $\alpha_s$ than that of the star-forming population that is clearly not present in the mass function of central galaxies (P12) or in low density environments (P10), for both of which environment-quenching can be neglected. Furthermore, since the value of the sSFR is observed to be larger at higher redshifts, the steepening of $\alpha_s$ due to a negative value of β will be faster at higher redshifts. The action of this additional mass-dependent quenching term λ' will therefore be more efficient at higher redshifts and as a consequence, will produce many passive low mass galaxies at high redshifts, which have not been observed.

Therefore, we conclude that an additional low-mass quenching term is not a viable solution to the steepening problem. The excess low mass galaxies produced by the steepening of the mass-function must be eliminated and not just quenched, to avoid producing too many low mass passive galaxies.

## 4  MERGING AND THE EFFECTIVE SSFR

As we demonstrated in P10 and P12, the combined effect of environment-quenching must be independent of the stellar mass, as required by the "separability" of the red fraction of galaxies in the SDSS. Thus environment quenching should have no effect on the faint end slope $\alpha_s$ of the population of star-forming galaxies. The mass-quenching process produces the exponential cut off in the mass function at M* and then, once it has been established, keeps the value of M* constant. The mass-quenching rate preserves the shape of the mass function and thus it should not change the value of $\alpha_s$.

Without environment-quenching or mass-quenching, and in the absence of any destruction of galaxies through merging, the change in the number density of star-forming galaxies due to the increase in their masses driven by *in situ* star formation will be given in terms of sSFR and β by Equation (6). As discussed in the previous section, $\alpha_s$ will become progressively more negative for a negative value of $\beta$, if $\alpha_s < -(1+\beta)$ as observationally indicated.

We now come to a key point in the paper. Suppose we start with a particular Schechter function of star-forming galaxies at some time *t*, as indicated by the red curve in the bottom panel of Figure 5, and further suppose that all these star-forming galaxies are increasing their masses through *in situ* star-formation that is described by some sSFR(*m*) function with $\beta \neq 0$, as shown by the green line in the top panel of Figure 5. This star-formation activity will clearly shift the mass function to higher masses and (since $\beta \neq 0$) steepen the faint end slope to produce the green mass-function in the lower panel.

We now imagine that merging will modify this green mass-function into the black one (also shown in the lower panel), by both destroying some galaxies (especially at low masses) and by increasing the masses of some galaxies (especially at high masses). The action of merging must of course preserve the total stellar mass of the population. Clearly, if merging is to solve the problem of constant $\alpha_s$, then this modified mass function (in black in the lower panel) must have the same faint end slope as the initial mass function that was shown in red.

Given this, we must then be able to find another, "virtual" or "effective", sSFR$_{eff}$ function that will produce the black mass function directly from the red one *without* any merging and with no mass-dependence, i.e. $\beta_{eff} = 0$, since we know from Section 3 that *any* sSFR function with $\beta = 0$ will preserve the faint end slope. This effective sSFR$_{eff}$ is shown as the black line in the top panel, and is by construction independent of mass. We stress that this sSFR$_{eff}$ is in no sense an observable quantity and will not have any validity for any individual galaxy, which is why we think of it as a virtual quantity. Indeed, its only validity is in artificially reproducing changes to the mass function of the population of (star-forming) galaxies and thereby greatly simplifying the analysis. In particular one should not think of the sSFR$_{eff}$ as the simple sum of the real sSFR and the specific mass increase due to mergers (sMMR) that we introduced above. This is because the action of the sSFR$_{eff}$ must reproduce also the change in number density due to merging in the real Universe, as we will discuss in the next Section.

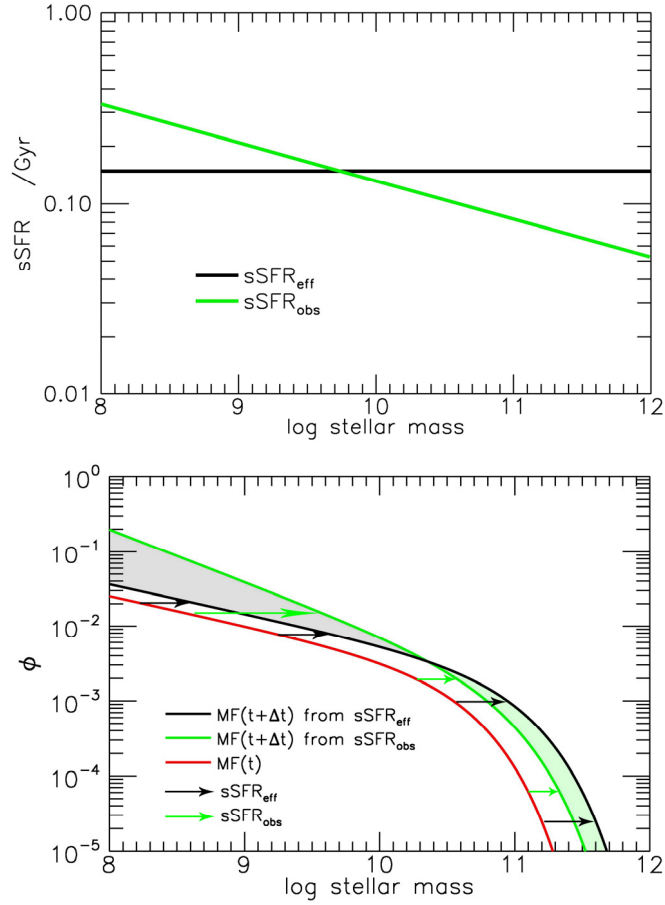

Figure 5. Schematic plots to show the equivalence of the evolution of the mass function of star-forming galaxies from the observed sSFR($m$) with merging, and from the virtual sSFR$_{eff}$ without merging. In the top panel, the green line is the observed sSFR($m$) with a negative β. In the lower panel, we start from the red mass function and produce the green one by applying this observed sSFR($m$) relation. This has a steeper faint-end slope because of the differential mass increase indicated by the green arrows. The action of merging will be to modify this green mass function into the black one, both by destroying some galaxies (especially at the low mass end) and by increasing the masses of others (especially at the high mass end). If merging is to keep $\alpha_s$ constant, then this black mass function must have the same faint end slope as the red one. Since merging preserves the total mass in the population, the light grey and light green shaded areas must contain equal mass. However, the black mass function can also be produced by applying a virtual effective sSFR$_{eff}$ with $\beta_{eff} = 0$ to the original red mass function, which will produce a shift in the mass function that is independent of mass (as indicated by the black arrows). The constant sSFR$_{eff}$ is also shown in the upper panel of the Figure. It should be emphasized that the sSFR$_{eff}$ is not an observable quantity for any galaxy, as discussed in the text.

Based on Equation (13) above, and setting aside the effects of quenching, the change of the mass function of star-forming galaxies as a result of this virtual sSFR$_{eff}$ will be given by

$$\frac{1}{\phi_{blue}}\frac{\partial \phi_{blue}}{\partial t} = -\text{sSFR}_{eff} \cdot (1+\alpha_s - \frac{m}{M^*} + \beta_{eff}) \tag{14}$$

where sSFR$_{eff}$ is by construction mass independent with $\beta_{eff} = 0$ and we have also used $\alpha = (1+\alpha_s) - m/M^*$.

Although galaxy mergers reduce galaxy number density, they conserve mass (neglecting star-formation during the merger itself). Therefore the total stellar mass increase during a short interval of time due to the real observed sSFR (represented by the region marked with green arrows in the bottom panel of Figure 5) must be equal to the total stellar mass increase driven by the virtual sSFR$_{eff}$ (represented by the region marked with black arrows in the same plot). We can therefore write

$$\int_0^\infty \left.\frac{d\phi_{blue}}{dt}\right|_{sSFR} m\,dm = \int_0^\infty \left.\frac{d\phi_{blue}}{dt}\right|_{sSFR_{eff}} m\,dm$$

$$\int_0^\infty -\phi_{blue} \cdot sSFR \cdot (1+\alpha_s - \frac{m}{M^*} + \beta) m\,dm = \int -\phi_{blue} \cdot sSFR_{eff} \cdot (1+\alpha_s - \frac{m}{M^*}) m\,dm \quad (15)$$

Inserting the definitions of sSFR($m$) from Equation (4) and of the Schechter function of $\phi_{blue}$ from Equation (1) into Equation (15), gives

$$k(1+\alpha_s+\beta)\Gamma(\alpha_s+\beta+2) - k\Gamma(\alpha_s+\beta+3) =$$
$$sSFR_{eff} \cdot (1+\alpha_s)\Gamma(\alpha_s+2) - sSFR_{eff} \cdot \Gamma(\alpha_s+3) \quad (16)$$

where $\Gamma$ is the gamma function defined in Equation (3). In principle, we should integrate the gamma function from the minimum mass of the galaxy ($M_{min}$) to infinity, i.e. use the incomplete gamma function $\Gamma(\alpha_s+\beta+2, M_{min}/M^*)$ instead of the complete gamma function $\Gamma(\alpha_s+\beta+2)$. However, for the observed $\alpha_s \sim -1.4$, $\beta \sim -0.1$ there is a negligible error in using the complete $\Gamma$ function since, for $M_{min} \sim 10^6 M_\odot$, then $\Gamma(\alpha_s+\beta+2, M_{min}/M^*)=1.762$ and $\Gamma(\alpha_s+\beta+2)=1.772$ while $\Gamma(\alpha_s+\beta+3, M_{min}/M^*)=0.886$ and $\Gamma(\alpha_s+\beta+3)= 0.886$. Thus we will simply use the complete gamma function in Equation (16) as a good approximation to the incomplete gamma function.

The Gamma function satisfies the functional equation:

$$\Gamma(x+1) = x\Gamma(x) \quad (17)$$

With Equation (17), Equation (16) can therefore be further simplified to:

$$sSFR_{eff}(t) = k(t)\frac{\Gamma(\alpha_s+\beta+2)}{\Gamma(\alpha_s+2)} \quad (18)$$

When $\beta$ is close to zero, then $\Gamma(\alpha_s+\beta+2) \sim \Gamma(\alpha_s+2)$ and SFR$_{eff} \sim k(t)$, where $k(t)$ is the value of sSFR($t$) at $m = M^*$. As illustrated in the top panel of Figure 5 when $m < M^*$, the virtual sSFR$_{eff}$ is less than the actual sSFR, i.e. the observed sSFR will produce more galaxies than are in the final mass function (produced by the sSFR$_{eff}$) and thus statistically the low mass galaxies with $m < M^*$ tend to be destroyed by mergers. The opposite is true for massive galaxies with $m > M^*$, where the observed sSFR does not produce enough galaxies. On average, these galaxies have accreted smaller galaxies and thereby increased their masses above what was produced by their in situ star-formation. This of course makes physical sense.

## 5 CONSTRAINTS ON MERGERS THROUGH THE DESTRUCTION OF GALAXIES

In this section we further develop the constraints on the required mergers so as to keep the faint-end slope $\alpha_s$ constant and demonstrate the underlying connections between $\alpha_s$, $\beta$ and the required amount of merging via the continuity equations and compare these with observations.

### 5.1 The required destruction and the mean sMMR of star-forming galaxies

First we introduce $\kappa(m)$ as the destruction rate of galaxies in mergers, i.e. it is the probability that a given galaxy of mass $m$ is destroyed by merging with another, more massive, galaxy. This has to be a function of mass if merging is to modify the faint-end slope. We will show below that $\kappa(m)$ must have the same mass-dependence as the observed sSFR. Analogous to the sSFR, we also now introduce the specific merger mass rate (sMMR) as the specific rate of mass increase of the surviving galaxy from the accretion of lower mass galaxies. Clearly for any surviving individual galaxy, the actual mass increase will be given by the sum sSFR+sMMR. As noted above, this is *not* the same as the sSFR$_{eff}$.

We define $x$ as the typical mass ratio of galaxies that merge and we will assume for simplicity that $x$ is independent of $m$. This assumption is likely to be true in the power law part of the mass function but may well fail in the region of the exponential cut-off. We will see below that neither $\kappa$ nor the sMMR depend strongly on $x$ and so the use of a single typical value, rather than the extended distribution that likely exists in reality, should not be a major issue.

Since the constraint on mergers via the destruction of galaxies is essentially based on the constancy of $\alpha_s$, i.e. on the power law part of the mass function, our analysis in this section focuses mainly on that part of the mass function. The region of the exponential cut off will, as discussed in P10, be dominated by the removal of star-forming galaxies in the mass-quenching process. We return to this point below.

The destruction of one galaxy of mass $m$ will be associated with the gain in mass of another galaxy at higher mass $xm$. This "local" mass conservation during mergers then requires the following,

$$\phi(m)\kappa(m)m = \phi(xm) \cdot \text{sMMR}(xm) \cdot xm \qquad (19)$$

The Schechter mass-function of the star-forming galaxies in units of the number density per log $m$ is given by

$$\phi_{blue}(m)\, d\log m = e^{-\frac{m}{M^*}} \left(\frac{m}{M^*}\right)^{\alpha_s+1} \phi^*_{blue} \ln 10\, d\log m \qquad (20)$$

so, in the power law part of the mass function, i.e. at $m \ll M^*$, we have

$$\phi_{blue}(xm) = \phi_{blue}(m)x^{\alpha_s+1} \qquad (21)$$

Inserting Equation (21) into (19) then gives

$$\kappa(m) = x^{\alpha_s+2} \cdot \text{sMMR}(xm) \qquad (22)$$

We then assume that κ and sMMR both have the analytical form of a power-law plus a constant

$$\kappa(m) = k_\kappa \left(\frac{m}{M^*}\right)^{\beta_\kappa} + c_\kappa \qquad (23)$$

$$\text{sMMR}(m) = k_m \left(\frac{m}{M^*}\right)^{\beta_m} + c_m \qquad (24)$$

where $k_k$, $k_m$, $c_k$ and $c_m$ are some mass independent, but possibly time-dependent, terms to be determined. Inserting Equations (23) and (24) into (22), we then require

$$\beta_\kappa = \beta_m \qquad (25)$$

$$c_\kappa = x^{\alpha_s+2} c_m \qquad (26)$$

$$k_\kappa = x^{\alpha_s+\beta_m+2} k_m \qquad (27)$$

Now turning to the change in the mass function of star-forming galaxies, we can write the continuity equation (see Equation 10 of P10) in terms of the sSFR, sMMR and κ as

$$\frac{1}{\phi_{blue}}\frac{\partial \phi_{blue}}{\partial t} = -\text{sSFR}\cdot(\alpha+\beta) - \kappa - \text{sMMR}\cdot(\alpha+\beta_{sMMR}) = -\text{sSFR}_{eff}\cdot\alpha \qquad (28)$$

where α is the local logarithmic slope of the mass function and is given by α = (1+α_s)–m/M* for a Schechter function. In the power law part of the mass function, we can drop the –m/M* term in α and Equation (28) is then simplified to

$$\frac{1}{\phi_{blue}}\frac{\partial \phi_{blue}}{\partial t} = -\text{sSFR}\cdot(1+\alpha_s+\beta) - \kappa - \text{sMMR}\cdot(1+\alpha_s+\beta_{sMMR}) = -\text{sSFR}_{eff}\cdot(1+\alpha_s)$$

(29)

where $\beta_{sMMR}$ is the local logarithmic slope of the sMMR-*m* relation and is given in terms of the $\beta_m$ introduced above as

$$\beta_{sMMR} = \frac{d\log \text{sMMR}}{d\log m} = \beta_m\left(1 - \frac{c_m}{\text{sMMR}}\right) \qquad (30)$$

Inserting Equations (23)-(25) and (30) into Equation (29) gives

$$\beta = \beta_\kappa = \beta_m \qquad (31)$$

$$k_m = k\frac{-(\alpha_s+\beta+1)}{x^{\alpha_s+\beta+2}+(\alpha_s+\beta+1)} \qquad (32)$$

$$c_m = k\frac{\Gamma(\alpha_s+\beta+2)}{\Gamma(\alpha_s+2)}\frac{\alpha_s+1}{x^{\alpha_s+2}+(\alpha_s+1)} \qquad (33)$$

Inserting Equations (32) and (33) into (26) and (27) gives

$$k_\kappa = k \frac{-(\alpha_s + \beta + 1)x^{\alpha_s+\beta+2}}{x^{\alpha_s+\beta+2} + (\alpha_s + \beta + 1)} \quad (34)$$

$$c_\kappa = k \frac{\Gamma(\alpha_s + \beta + 2)}{\Gamma(\alpha_s + 2)} \frac{(\alpha_s + 1)x^{\alpha_s+2}}{x^{\alpha_s+2} + (\alpha_s + 1)} \quad (35)$$

Putting Equations (31)-(35) into Equations (23) and (24) finally gives

$$\kappa(m) = \frac{-(\alpha_s + \beta + 1)x^{\alpha_s+\beta+2}}{x^{\alpha_s+\beta+2} + (\alpha_s + \beta + 1)} \cdot \text{sSFR}(m) + \frac{(\alpha_s + 1)x^{\alpha_s+2}}{x^{\alpha_s+2} + (\alpha_s + 1)} \cdot \text{sSFR}_{eff} \quad (36)$$

$$\text{sMMR}(m) = \frac{-(\alpha_s + \beta + 1)}{x^{\alpha_s+\beta+2} + (\alpha_s + \beta + 1)} \cdot \text{sSFR}(m) + \frac{\alpha_s + 1}{x^{\alpha_s+2} + (\alpha_s + 1)} \cdot \text{sSFR}_{eff} \quad (37)$$

Equations (36) and (37) are the exact analytical solutions of Equation (29) in the power law part of the mass function. In fact, it is straightforward to show that if β is close to zero, then these are also good approximations in the exponential cut-off region, i.e. they are approximate solutions to the more general Equation (28).

It is clear from Equations (36) and (37) that, for equal mass mergers, i.e. $x \sim 1$, then κ=sMMR. This is expected since equal mass mergers will halve the numbers of the merger progenitors and double the masses of the surviving merged galaxies, so the average sMMR across the population will equal the fraction of galaxies being destroyed.

As noted at the start of the paper, a key question in understanding the assembly of the stellar mass of galaxies is to compare the sSFR and sMMR averaged across the star-forming population at a particular mass. In Figure 6 we show the ratio of sMMR/sSFR as a function of β for different values of $\alpha_s$ and $x$.

Several points can be noticed from this Figure. First there is a strong dependence of the required sMMR/sSFR ratio on the value of β. More negative value of β make it increasingly hard to keep $\alpha_s$ constant. Indeed, it becomes impossible to control $\alpha_s$ when $\alpha_s+\beta+2 \leqslant 0$, i.e. for $\beta \leqslant -(\alpha_s+2)$. Since we know observationally that $\alpha_s \sim -1.4$, it is hard to see how "true" values of $\beta \sim -0.5$ could possibly be consistent with a constant observed $\alpha_s$.

Second, for a given $\alpha_s$, the sMMR/sSFR ratios for different mass ratios $x$ converge to a lower asymptote as $x$ becomes very large. When $x$ is very large, the more massive galaxies in the merger events can consume many low mass galaxies with little corresponding increase in their own stellar mass, and thus it requires a relatively smaller sMMR to keep $\alpha_s$ constant.

Third, with our preferred values of $\alpha_s \sim -1.4$ and $\beta \sim -0.1$, the required sMMR/sSFR ratio is around 0.1. This does not depend strongly on $x$ or $\alpha_s$, as indicated by the red box enclosing possible solutions. We conclude that sMMR ~ 0.1sSFR, which implies that, on average across the population, about 90% of the stellar mass is produced in *in situ* star-formation as opposed to 10% that is brought in to the galaxy by mergers. Clearly this is only a rough estimate - the box on Figure 6 spans a factor of three range in sMMR/sSFR. However, we can conclude from this analysis that stellar mass assembly through merging is probably a significant but not dominant process in the buildup of the stellar mass of a typical star-forming galaxy.

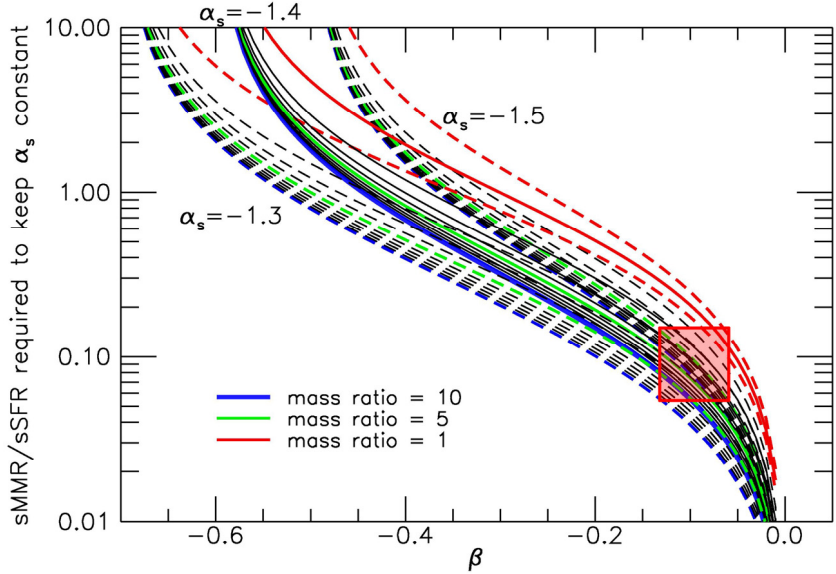

Figure 6. The ratio of sMMR/sSFR as a function of β for different values of $\alpha_s$ and $x$. The curves for given $\alpha_s$ quickly converge towards the asymptote when the mass ratio $x$ becomes very large (blue lines). More negative values of β require higher sMMR and make it increasingly difficult to control $\alpha_s$. It becomes impossible for $\alpha_s+\beta+2<0$, i.e. for $\beta< -(\alpha_s+2)$. For the observed values of $\alpha_s \sim$ -1.4 and β ~ -0.1, as indicated by the red box, it implies sMMR ~ 0.1sSFR, i.e. on average across the whole population, mergers bring about 10% of the stellar mass that is assembled through star-formation.

## 5.2 Merger rates

Following usual observational practice, we will define the galaxy merger rate $R(m)$ as the total number of galaxies that merge into galaxies of a given stellar mass $m$, divided by the total number of such galaxies. In other words, it is the probability per unit time that a given galaxy of mass $m$ will have accreted a lower mass galaxy. This will clearly be related to the *sMMR(m)* through $x$, the typical mass ratio, as follows,

$$R(m) = x \cdot \text{sMMR}(m) \qquad (38)$$

To see if this merger rate is plausible observationally, we can consider two cases where the merging is primarily in major mergers ($1 \leqslant x \leqslant 4$, i.e. with $\langle x \rangle \sim 2$) and in observationally recognizable minor mergers ($4 < x \leqslant 10$, with a plausible $\langle x \rangle \sim 7$). With sMMR ~ 0.1 sSFR for $\alpha_s \sim$ -1.4 and β ~ -0.1 from Figure 6, we would expect a merger rate in the sky of order 0.2 sSFR if the dominant mass increase is due to major mergers, or of order 0.7 sSFR if it is due to minor mergers. If both regimes were important, each would be correspondingly reduced.

Encouragingly, despite the uncertainties in our own prediction and the considerable observational uncertainties, these predicted merger rates agree rather well with observational estimates from the literature, as shown in Figure 7.

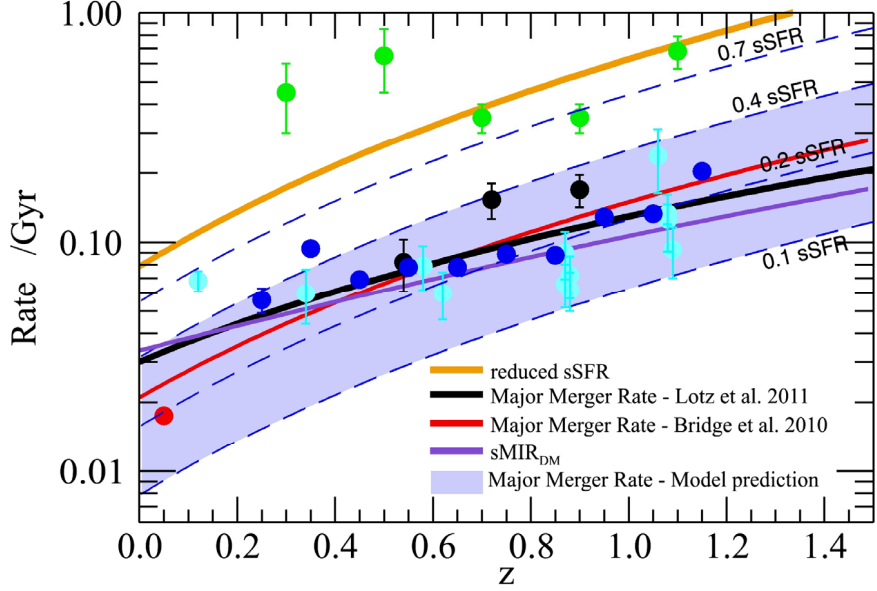

Figure 7. The observed reduced sSFR at $m = 10^{10}M_\odot$ (brown line) and observed merger rates from the literature, as a function of redshift. The black line and red line show the best fit of the observed major merger rates per galaxy from Lotz et al. (2011, $\Re_{pairs,PLE}$ in their Table 4) and Bridge et al. (2010), respectively. The dashed blue lines show the predicted average major merge rate of 0.2sSFR for $<x> \sim 2$ and the predicted minor merger rate of 0.7sSFR for $<x> \sim 7$. The light blue shaded regions between 0.1sSFR and 0.4sSFR show the predicted possible range of major merger rates for $1 \leqslant x \leqslant 4$ from the model, as discussed in the text. The purple line shows the median specific Mass Increase Rate (sMIR) of typical dark matter haloes determined from Faucher-Giguere et al. (2011), for reference. The data points are taken from the reconciled major merger rates from recent studies and listed in the Table 1 & 2 of Lotz et al. (2011), i.e. Patton & Atfield (2008) in red; Lin et al. (2008) in cyan; Kartaltepe et al. (2007) in blue; de Ravel et al. (2009) in black. The green points represent an estimate of all mergers (including minor ones) from the morphological $G - M_{20}$ analysis of Lotz et al. (2008).

### 5.3 Caveats on the derived κ and sMMR

To derive the analytic forms of κ and sMMR from the constraint of a constant $\alpha_s$ with cosmic time coupled with a negative value of β, we have implicitly made two assumptions, which we here address.

The first assumption was when, in Section 4, we introduced the effective specific mass increase sSFR$_{eff}$ to represent the combined effects of the (observed) sSFR, the destruction by merging, κ, and the mass increase due to merging, sMMR. We assumed that the sSFR$_{eff}$ is independent of mass, i.e. has β$_{eff}$

= 0. This is clearly required in the power-law regime below M* in order to keep the resulting $\alpha_S$ constant, but what about the region above M*? This region is in a sense "hidden" because of the action of mass-quenching which imposes an exponential cut-off in the mass-function of star-forming galaxies, and there is the possibility that the sSFR$_{eff}$ could deviate from constancy above M*. However, in our earlier papers (P10 and P12) we showed that we could obtain an excellent fit to the observed mass-functions of different populations in SDSS by assuming $\beta \sim 0$ (i.e. a constant sSFR) across the whole mass range, and neglecting the effect of mergers. This is equivalent to using an effective sSFR (with mergers) with $\beta_{eff} \sim 0$ across the whole mass range, and suggests that the assumption of constant sSFR$_{eff}$ is likely to be a good approximation to reality. Put another way, a value of $\beta_{eff}$ that was not zero above M* would mess up the simple relations between the mass functions of star-forming and passive galaxies that were noted in P10, specifically $\Delta\alpha_s = 1+\beta$. Indeed, the mechanism via merging that we introduced here to produce $\beta_{eff}$ close to zero offers an explanation as to why the faint end slopes of the star-forming and passive populations in P10 differ by almost exactly unity even though the observed $\beta$ is clearly not zero. We return to this in Section 6.3 below.

The second assumption was that the mass of destroyed star-forming galaxies ends up in other star-forming galaxies. In other words, we neglected the possibility that star-forming galaxies might merge into passive galaxies that stay passive, into other star-forming galaxies that become passive because of the merger, or, in the other direction, that they merged into passive galaxies that started to actively form stars again in a rejuvenation process. In addition, star-forming galaxies could grow by accreting smaller passive galaxies. These four processes would each perturb the equality in Equation (15), the first two by decreasing the integral on the LHS of Equation (15), and the second two by increasing it. Ignoring these effects is equivalent to assuming that these perturbations to the equality cancel out. It should be clear that the problem of keeping $\alpha_s$ constant that has been the central theme of the paper is most acute at high redshifts where the cosmic sSFR is high. At these redshifts, the number of passive galaxies at low masses is small, and it is therefore unlikely that passive galaxies seriously perturb the integrals at low masses. At higher masses, the situation is less clear, but the arguments in the previous paragraph would then apply.

Currently there is not much observational evidence as to whether the lack of evolution in $\alpha_s$ is maintained across different environments, although we note that the present-day $\alpha_s$ does not seem to depend much on environment (P10). Since the sSFR and $\beta$ are also more or less independent of environment (P10, P12), then we might expect the required merger rate from the current analysis should also be independent of environment. This may seem to contradict the fact that the major merger rate is observed to be higher in dense regions (Kampczyk et al. 2013, also as argued in P10). However, the mergers concerned in this paper are mainly high mass ratio minor mergers and it is not clear that whether the minor merger rate is dependent on environment or not. In fact the high mass ratio minor mergers behave similar to the regular accretion onto the dark matter haloes. Since the sSFR of the star-forming galaxies is found to be independent of the environment (P10, P12), given the tight relation between the sSFR and the specific Mass Increase Rate (sMIR) of typical dark matter haloes (Lilly et al. 2013), one would hence expect the sMIR (from minor mergers) to be also quasi-independent of environment.

# 6 DISCUSSION

## 6.1 The connection between the sMMR and the assembly of dark matter halos

In Section 5.1 we found that with an sMMR ~ 0.1 sSFR, merging will be able to keep $\alpha_s$ constant over time for $\beta \sim -0.1$. This implies a major merger rate of order ~ 0.2 sSFR. We showed in Section 5.2 that this is quite consistent with observational estimates of the major merger rates.

Why should the merger rate know anything about the cosmic evolution of the sSFR of Main Sequence galaxies? As mentioned in P10 and discussed in detail in Lilly et al (2013) and Peng et al (2014), there are clear similarities between the time evolution of the sSFR of star-forming main sequence galaxies and the time evolution of the specific Mass Increase Rate (sMIR) of typical dark matter haloes. Both quantities are also only weakly dependent on mass, with $\beta_{DM} \sim +0.1$ and $\beta \sim -0.1$.

These may be explained in terms of simple gas-regulated model of star-formation in galaxies (Lilly et al 2013; Peng et al 2014). The growth of haloes will also be associated with the accretion of already existing galaxies, and so we might well expect the merging rate of galaxies (at fixed stellar and/or halo mass) to also follow the overall sMIR of the haloes. Thus, on quite general grounds we could expect a link between the star-formation rates in galaxies and the addition of mass through merging of pre-existing galaxies.

This illustrates the power of our analytic and phenomenological approach to galaxy evolution in revealing interconnections between different aspects of galaxy evolution. The slope of the mass function, the slope of the main sequence sSFR-$m$ relation and the merger rate may at first sight appear as independent/uncorrelated properties of star-forming galaxies. Yet, in this paper we uncover that they are intimately interconnected, and they must satisfy stringent mutual relations which are demanded by the observed evolution of these quantities. We would argue that the remarkable and surprising constancy of the faint end slope of the star-forming mass function arises (assuming the true star-forming $\beta$ is non-zero) from a deep link between the sSFR of galaxies and the growth of their dark matter haloes.

## 6.2 The intrinsic scatter in the SFR - stellar mass relation

All our analyses above are done based on the average SFR-stellar mass relation for the star-forming galaxy population. Could the intrinsic scatter in the observed SFR-stellar mass relation have any influence of our conclusions, by modifying the evolution of $\alpha_s$? The intrinsic scatter in the SFR is likely to be due to up and down fluctuations of the SFR around the mean value, given the stochastic nature of the star formation process (Salmi et al. 2012). For instance, the gas inflow of individual galaxies is likely to be clumpy within short timescale. Therefore, the random scatter in the SFR-stellar mass relation is unlikely to have any significant effect on the evolution of $\alpha_s$.

To investigate this issue quantitatively, we add a random scatter to the average SFR-mass relation in the simple simulation as constructed in P10, with a standard deviation of the SFR distribution (at any given stellar mass) $\sigma = 0.2$, which is about the observed value of $\sigma \sim 0.19$ in Sargent et al (2012). In particular, we test different timescales of the fluctuation of the SFR around the mean, parameterized by $\tau_{fluc}$. For instance, if we assume $\tau_{fluc} = 1$Gyr, it means that for a given star-forming galaxy, we deviate its SFR from the mean value by a fixed random factor for one billion years and we then change this factor randomly every one billion years. Different values of $\tau_{fluc}$ will test the influence of scatter in the SFR-mass relation on the $\alpha_s$ evolution for both short-term and long-term deviation of the SFR from the mean.

As in P10, we start the simulation from z ~ 10 with a primordial logarithmic star-forming mass function that is a power law with slope equivalent to a Schechter faint end slope of $\alpha_s = -1.4$. To clearly demonstrate the effect of the scatter in the SFR on the evolution of $\alpha_s$, we let $\beta = 0$; assume there is no merging and no environment-quenching; and keep only mass-quenching.

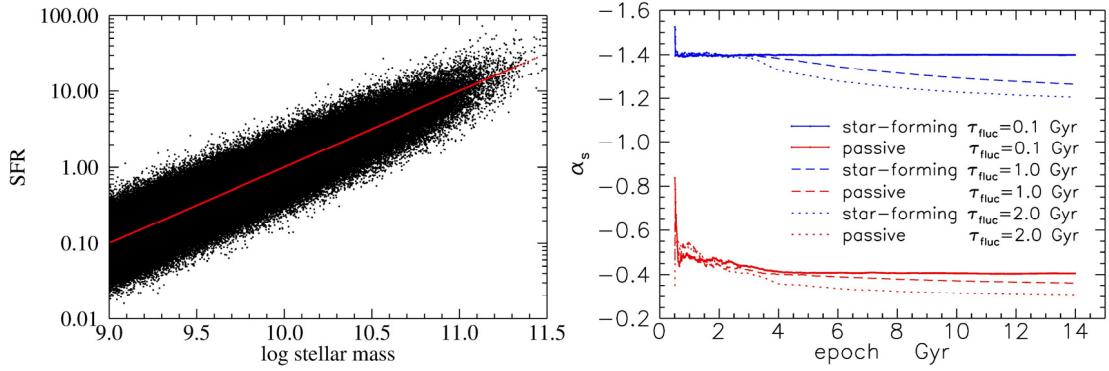

Figure 8. Left panel: The SFR-stellar mass relation at z ~ 0 with a random scatter of $\sigma = 0.2$ in the simple simulation described in the text. The red line shows the average SFR-stellar mass relation. Right panel: Associated evolution of the faint end slope of the mass functions of star-forming galaxies and mass-quenched passive galaxies for different values of the fluctuation timescale $\tau_{fluc}$ as described in the text. For short $\tau_{fluc}$, the $\alpha_s$ of the star-forming mass function is maintained at the initial value of −1.4 at all epochs and the $\alpha_s$ of the mass-quenched passive mass function quickly stabilizes to the expected value of −0.4, as predicted in P10. A longer $\tau_{fluc}$ makes the $\alpha_s$ of the star-forming and passive mass functions slightly less negative. Even for $\tau_{fluc}$=2 Gyr, the $\alpha_s$ of the star-forming mass function changes by a maximum amount of ~0.2, which is much smaller than the change of $\alpha_s$ due to a negative value of $\beta$ shown in the right panel of Figure 3. Therefore the random scatter in the SFR-mass relation has very little influence on our conclusion that the $\alpha_s$ of the star-forming mass function will quickly steepen for a negative value of $\beta$.

The SFR-stellar mass relation in the simulation at z ~ 0 is plotted in the left panel in Figure 8 as the black dots. The red line in the same panel shows the average SFR-stellar mass relation. In the right panel, we plot the associated evolution of the faint end slope of the mass function for star-formation galaxies (blue lines) and mass-quenched passive galaxies (red lines) as a function of cosmic epoch, for different values of the fluctuation timescale $\tau_{fluc}$.

For short fluctuation timescale of $\tau_{fluc} \leq 0.1$Gyr, the faint end slope of the star-forming mass function is maintained at the initial value of −1.4 (as a consequence of $\beta = 0$) at all epochs. The faint end slope of the passive mass function quickly stabilizes to the expected value of −0.4, as predicted in P10 that the faint end slope of the star-forming mass function and of the mass-quenched passive mass function should differ by one. This is naturally expected since the fluctuation of the SFR on short timescales will be averaged out and it hence has negligible influence on the evolution of the faint end slope of the mass function.

A longer $\tau_{fluc}$ evidently makes the $\alpha_s$ of the star-forming and passive mass functions slightly less negative. However, even for $\tau_{fluc}$=2 Gyr, it changes the $\alpha_s$ of the star-forming mass function by a maximum amount of ~0.2, which is much smaller than the change of $\alpha_s$ due to a negative value of β as shown in the right panel of Figure 3. Given the stochastic nature of the star formation process, $\tau_{fluc}$ is expected to be short and it is usually seen to be small in hydrodynamic simulations (e.g. Dekel et al. 2013). Therefore, the intrinsic scatter (in additional to the scatter due to observation errors and uncertainties) in the SFR-mass relation should have only a small influence on the evolution of the faint end slope of the mass function for both star-forming and passive galaxies. It has very little influence on our conclusion that the faint end slope of the star-forming mass function will quickly steepen for a negative value of β.

### 6.3 Modification to the P10 formalism

In principle, the P10 formalism already took a non-zero β into account and used β as an input parameter in most of the analysis and equations. However, as discussed in Section 3, a negative β will not only complicate the analytical solutions to these equations in P10, but also cause the $\alpha_s$ steepening problem and destroy the Schechter form of the mass function. We have argued above that this problem can be solved by destroying the low mass galaxies and (minor-) merging them into more massive galaxies. When applying this additional process to the P10 formalism, we simply need to substitute the observed sSFR (with its negative β) by the sSFR$_{eff}$ (with $β_{eff}$= 0) to account for the combined effects of the observed sSFR, sMMR and the destruction rate κ. Then all the analysis and results in P10 and P12 will remain the same. Therefore, this has in fact further simplified the P10 formalism, since in several cases we had to *assume* β=0 in order to find a simple analytic solution to the differential equations. This simplification may be justified by the current analysis. Examples include when we derive the analytic form of $\varepsilon_m$ (Equation 8 in P12). In our future work we may simply use sSFR$_{eff}$ (with $β_{eff}$=0) in all the continuity equations to replace the observed sSFR (with negative β).

It is also clear from Equation (18) that if the observed value of β is close to zero, sSFR$_{eff}$ ≈ $k$ = sSFR(M*). Therefore, if the observed value of β is modest negative, it is convenient to use the observed value of sSFR at $m$ = M* as an approximation to the sSFR$_{eff}$ of the star-forming galaxy population.

### 7. SUMMARY

Following the simple analytical framework of galaxy evolution established in P10 and P12, in this paper we explore the analytic consequence of the striking observed constancy of the faint end slope $\alpha_s$ of the star-forming mass function with epoch up to at least z~2, through the continuity approach. If the logarithmic slope β of the sSFR-mass relation is negative, then $\alpha_s$ should have quickly diverged, due to the differential mass increase of galaxies along the galaxy star-forming main sequence. This mass-dependent differential mass increase will also quickly destroy the Schechter shape of the mass function.

This problem can be solved by destroying the low mass galaxies by merging them into more massive galaxies. Such merging will help to reduce the number density of the low mass galaxies and enhance the mass increase rate (i.e. sSFR) of more massive galaxies, both of which act to counter the steepening of the mass function. In principle, this can keep $\alpha_s$ constant for a negative value of $\beta$ for suitable merger rates.

To quantify this process, we introduce $\kappa$ as the destruction rate of the galaxies by mergers and sMMR as the specific rate of mass added to a given galaxy through mergers. The observed mass function (with a constant $\alpha_s$) at given epoch is then given by the joint action of the observed sSFR, sMMR and $\kappa$, plus any additional quenching processes. We then introduce a useful quantity $sSFR_{eff}$ as the effective specific mass increase rate to describe the combined effects of sSFR, sMMR and $\kappa$. The observed mass function evolution then is simply determined by $sSFR_{eff}$ alone plus any additional quenching processes. This $sSFR_{eff}$ is a virtual quantity that is not directly observable and which has no physical significance whatsoever for *individual* galaxies. Rather, it is useful to describe the changes in the galaxy *population*.

The constancy of $\alpha_s$ with epoch requires that the $sSFR_{eff}$ is independent of mass in the power law part of the mass function at $m<M^*$ and we also assume (with some justification) that this is the case above $M^*$ also. This then allows us to derive a very simple analytic form of the required $sSFR_{eff}$, in terms of the observed sSFR and $\beta$. Since mergers conserve mass, we also derive through continuity equations the analytic form of sMMR and $\kappa$. For a modest negative value of $\beta \sim -0.1$, it requires a sMMR $\sim$ 0.1sSFR to keep $\alpha_s$ constant with epoch, i.e. when averaged across the whole star-forming galaxy population, the mass accreted via mergers is only 10% of the mass growth via star-formation. This implies a merger rate of order $0.1x$ sSFR, where $x$ is the mass ratio between the merging galaxies pair. Following the usual definition of major merger with a mass ratio $x$ between 1:1 and 1:4, this then predicts a merger rate of R $\sim$ 0.2sSFR by using an average mass ratio of $<x> \sim 2$ for major mergers. Remarkably, the predicted major merger rate shows excellent agreement with the observed major merger rates from $z \sim 0$ up to at least $z \sim 1.5$.

A more negative value of $\beta$ requires a higher sMMR and merger rate. But it becomes increasingly difficult to control $\alpha_s$, and impossible for $\alpha_s+\beta+2 < 0$, i.e. for $\beta < -(\alpha_s+2)$. We regard this as a strong argument against a negative slope $\beta$ of the sSFR-$m$ relation being very much steeper than $\beta \sim -0.2$.

The close analytic link between the build-up of stars through in situ star-formation (sSFR) and through the accretion of previously made stars (sMMR), i.e. sMMR $\sim$ 0.1sSFR, that is evidently required to keep $\alpha_s$ constant probably arises because both quantities are ultimately driven by the build-up of dark matter haloes. This illustrates one of the powerful outcomes of our analytic approach to galaxy evolution: a deep physical connection between two quantities, sSFR and sMMR, produces a remarkable observational symmetry in the Universe, i.e. a constant faint-end slope of the star-forming galaxy population, $\alpha_s$.

These new findings have further developed the analytical framework established in P10 and P12 by demonstrating the important underlying connections between $\alpha_s$, $\beta$ and merging. In fact this analysis simplifies the framework by providing a justification for using a mass-independent $sSFR_{eff}$ (with $\beta_{eff}=0$) to account for the combined effects of the observed sSFR (with non-zero negative $\beta$), mass increase due to merging sMMR, and the destruction of galaxies described by $\kappa$, even though the observed sSFR is (weakly) mass-dependent.

**ACKNOWLEDGEMENTS**

We gratefully acknowledge the anonymous referee for comments and criticisms that have improved the paper. This research has been supported by the Swiss National Science Foundation.

**APPENDIX**

Alternatively, the sSFR$_{eff}$ can be derived in the following way. In addition to Equation (15), the total stellar mass increase per unit time can be also written as

$$\begin{aligned}\frac{dM}{dt} &= \int_0^\infty \frac{dm}{dt} \phi_{blue}\, dm \\ &= \int_0^\infty mk\left(\frac{m}{M^*}\right)^\beta \phi_{blue}\, dm \\ &= k\phi^* M^* \Gamma(\alpha_s + \beta + 2)\end{aligned} \quad (39)$$

While the total stellar mass of the star-forming galaxy population at a given epoch is given by

$$\begin{aligned}M &= \int_0^\infty m\, \phi_{blue}\, dm \\ &= \phi^* M^* \Gamma(\alpha_s + 2)\end{aligned} \quad (40)$$

Therefore the average sSFR of the star-forming galaxy population, which will be the sSFR$_{eff}$, since this is the same for all galaxies, is given by

$$\begin{aligned}\text{sSFR}_{eff} &= \frac{1}{M}\frac{dM}{dt} \\ &= k\frac{\Gamma(\alpha_s + \beta + 2)}{\Gamma(\alpha_s + 2)}\end{aligned} \quad (41)$$

It might be unintuitive to see why Equation (15) is equal to Equation (39). First, it is straightforward to show the auxiliary equation that

$$\int_0^\infty \text{sSFR} \cdot (1+\alpha+\beta)\phi_{blue} m\, dm = \int_0^\infty \text{sSFR} \cdot \left(2+\alpha_s - \frac{m}{M^*}+\beta\right)\phi_{blue} m\, dm = 0 \quad (42)$$

Reforming the above equation, gives

$$\begin{aligned}\int_0^\infty \text{sSFR} \cdot \phi_{blue} m\, dm &= -\int_0^\infty \text{sSFR} \cdot (\alpha+\beta)\phi_{blue} m\, dm \\ \int_0^\infty \frac{dm}{dt}\phi_{blue}\, dm &= \int_0^\infty \frac{d\phi_{blue}}{dt} m\, dm\end{aligned} \quad (43)$$